# THE 59th KARL JANSKY LECTURE: DISCOVERING THE RADIO UNIVERSE

K.I. Kellermann

*National Radio Astronomy Observatory, 520 Edgemont Road,
Charlottesville, VA 22903, USA.
E-mail: kkellerm@nrao.edu*

**Abstract:** The NRAO/AUI 59th Karl Jansky Lecture was presented on 24 October 2024, 22 November, 2024, and 4 December 2024 in Charlottesville, Virginia, Socorro, New Mexico, and Green Bank, West Virginia, respectively. The lectures covered the circumstances of the author's start in radio astronomy, the demographics of radio astronomers, discussions of the outstanding, mostly serendipitous, discoveries made by radio astronomers over the past century, and concluded with reflections on the prospects for further new discoveries.

## 1 INTRODUCTION

I think I have attended almost every one of the previous 58 Jansky Lectures, starting with John Bolton, who gave the first Jansky Lecture back in 1966. Bolton was my PhD supervisor at Caltech and it was he who got me started in radio astronomy. But it sort of happened by accident and I only became a radio astronomer because I misunderstood a question that he asked me during a very brief interview and so I accidently gave him the answer that got me the job.

I did not go to Caltech planning to study astronomy. In fact, I had no idea that one could have a career as an astronomer. I was actually planning to study physics. I had an assistantship which required that I work for some faculty member for 15 hours a week, for which I would be paid $1800/yr. I was given two weeks to find something interesting and a professor that was willing to take me on. I went around and talked with various professors who kindly took the time to explain their research projects to me. I talked to professors working in nuclear physics, atomic physics, low temperature physics, solid state physics, etc. They were all very cordial, but they asked a lot of questions about courses I had taken and my grades, especially in mathematics. However, I did not find anything that sounded interesting, and I do not think they were interested in me either.

When my two weeks were running out, the head of the Physics Department, suggested that I go talk with Professor Bolton who was starting a new radio astronomy project. That sounded like it might be interesting. So, I went to Bolton's office. I saw his secretary sitting there and said, "I would like to see Dr. Bolton." She glared at me and said "That's Mr. Bolton over there." Bolton turned around and looks at me and says, "What do you want?" Everyone else had at least treated me cordially, but I had clearly interrupted whatever it was that Bolton was doing. And I was sort of confused Mr., Prof., Dr., thinking to myself "This is Caltech; not Podunk U. How can a Caltech professor not have a PhD." So I mumbled something about maybe being interested in finding out about radio astronomy or something. Then he asked me just two simple questions. He was not interested in what courses I had taken or whether I could solve partial differential equations, but he asked me what I knew about electronics, and I told him that I was a radio amateur; I could read circuit diagrams and I knew how to use a soldering iron and build equipment. That was a good answer.

Then he asked me the key question which I completely misunderstood, but because I misunderstood what he was asking me, I accidently gave him the answer he was looking for. He asked me how I felt about heights. Well, I knew the observatory he was building was at 4,000 feet elevation. I had spent the previous summer with a high school friend camping in the Rockies and the Sierras and hiking at 10, 12, even 14 thousand feet.

So I confidently told him, "Oh heights don't bother me at all." But that is not what he meant about heights. If he had known that I am actually kind of scared of heights (Figure 1) that would have been the end rather than the beginning of my career in radio astronomy. But there was no further discussion; he just told me I should take an empty desk in the next room. Wait a minute, I thought. "I was just inquiring. I'm not sure I want to do this. I need to think this over." Also, I was thinking to myself, "I'm not sure I want to work for this guy." But then I realized that my 2 weeks was almost up, I had to do something to earn my $1800 so I figured "OK. I'll do this radio astronomy thing for the first semester, that will give me time to find something better."

Well, that was 65 years ago! The past 65 years has really been a fantastic time for radio astronomy. Most of the cosmic phenomena we study today, stuff that we read about in the newspapers or on TV, were unknown back then. Things like quasars, pulsars, the cosmic background radiation, interstellar masers, gravitational lenses, black holes, dark matter, extrasolar planets, and lots of other things have all been discovered since I began to work in radio astronomy. Unfortunately, or at least unfortunately for me, I had nothing to do with any of these discoveries, but I did get to know most of the major players—that is, the scientists who made all these discoveries and who later shared their experiences with me.

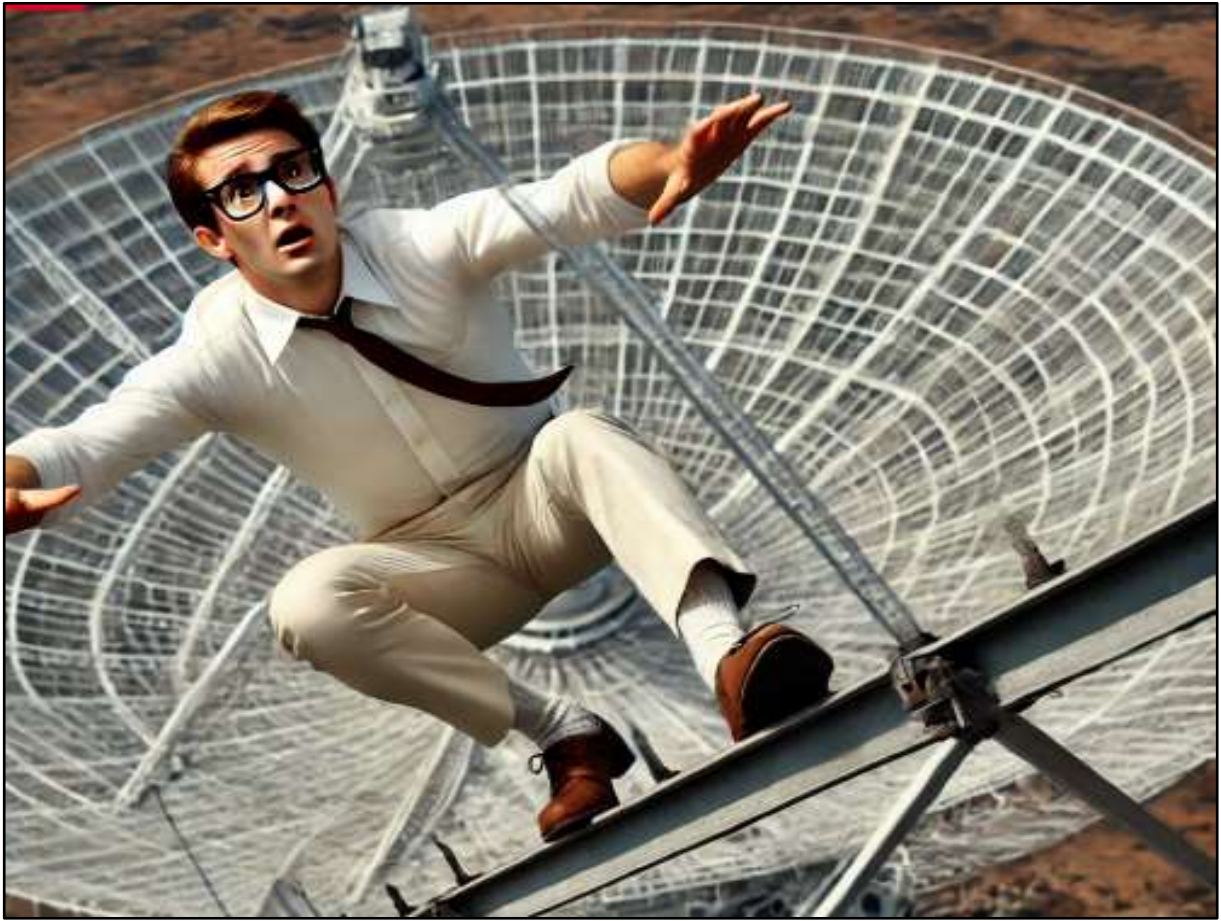

Figure 1: The author panicked when he realized that when Bolton asked about heights, he meant climbing on antennas not being at high altitudes.

Most of these discoveries were accidental or serendipitous. The scientists involved were mostly looking for something else, or in many cases they had a new instrument and were just looking, and often they accidently discovered something new. Generally, theory played no role in these discoveries, and in some cases incorrect theory pointed in a wrong direction and actually delayed a discovery. Mostly, the men and women involved were young, generally in their 20s and 30s, and some were still graduate students when they made their big discovery. Mostly these discoveries were the result of using new technologies or new techniques. Things like interferometry, aperture synthesis, digital signal processing, the ability to build large very precise radio telescopes, and of course the development of very sensitive receivers, were all critical in bringing about new astronomical discoveries.

In 1972 I had an undergraduate summer student named Stephen Chu. Chu went on to win a Nobel Prize for "Using lasers to cool and trap atoms." He later became Secretary of Energy under President Obama. I cannot even pretend to understand what he did, so I cannot claim any credit for his success, except to think perhaps he may have been influenced by what he saw happening at NRAO in radio astronomy which led him to later write:

> If you are the first person to look under a rock with a new set of tools, you don't even have to be that smart to discover something new (Chu, 2015).

Indeed, to a large extent it has been these new tools that has driven the history of radio astronomy. Although I think the people who made these pioneering discoveries were actually pretty smart. They were mostly engineers and physicists, very few astronomers, and in many cases, like John Bolton, they had only minimal academic training.

As you might expect, a lot of these discoveries were made in university laboratories or observatories, but some came from work that was done in industrial or military laboratories. And then, of course, there was Grote Reber, who didn't have any institutional affiliation and used his own personal funds to build the world's first radio telescope—essentially in his own backyard.

## 2　KARL JANSKY AND GROTE REBER OPEN A NEW WINDOW TO THE UNIVERSE

I never met Karl Jansky. Sadly, he died when I was still in middle school; but it has been a real honor to get to know his family—including his late wife Alice, as well as Karl's son and daughter, David and Ann Moreau (Figure 2).

Karl Jansky did not set out to discover cosmic radio emission. He was actually working for the telephone company. Back in the 1930s, there weren't any satellites for telecommunications. Instead, AT&T was using short wave radio for their transatlantic phone connections. Jansky's job was to find out what was causing disruption to these radio links. In order to find the source of interference, Jansky built a directional antenna which operated at 15 meter wavelength (Figure 3). His antenna spun around once every 20 minutes. His kids played on it and called it their merry-

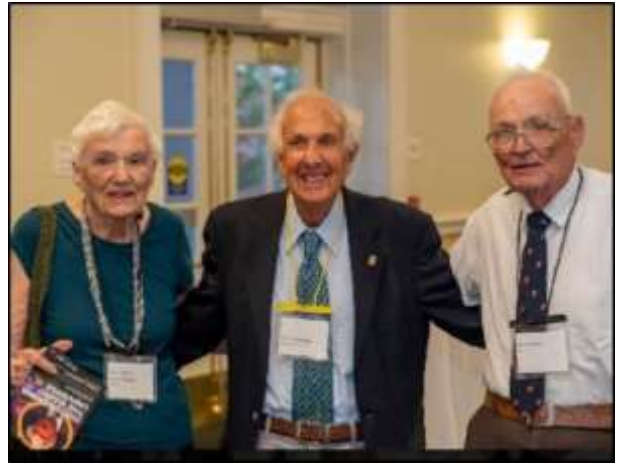

Figure 2 (left to right): Ann Moreau Jansky, the author, and David Jansky at the Charlottesville Jansky Lecture, 24 October 2024 (Kellermann Collection).

go-round. Radio noise was recorded on long rolls of chart paper.

During the winter of 1931/1932, Jansky noticed that the noise peaked three times per hour, and that the maximum noise he was receiving came when his antenna was pointed toward the Sun (Figure 4). Jansky had no background in astronomy and didn't appreciate that in the winter the Sun happens to lie in nearly the same direction as the center of the Milky Way.

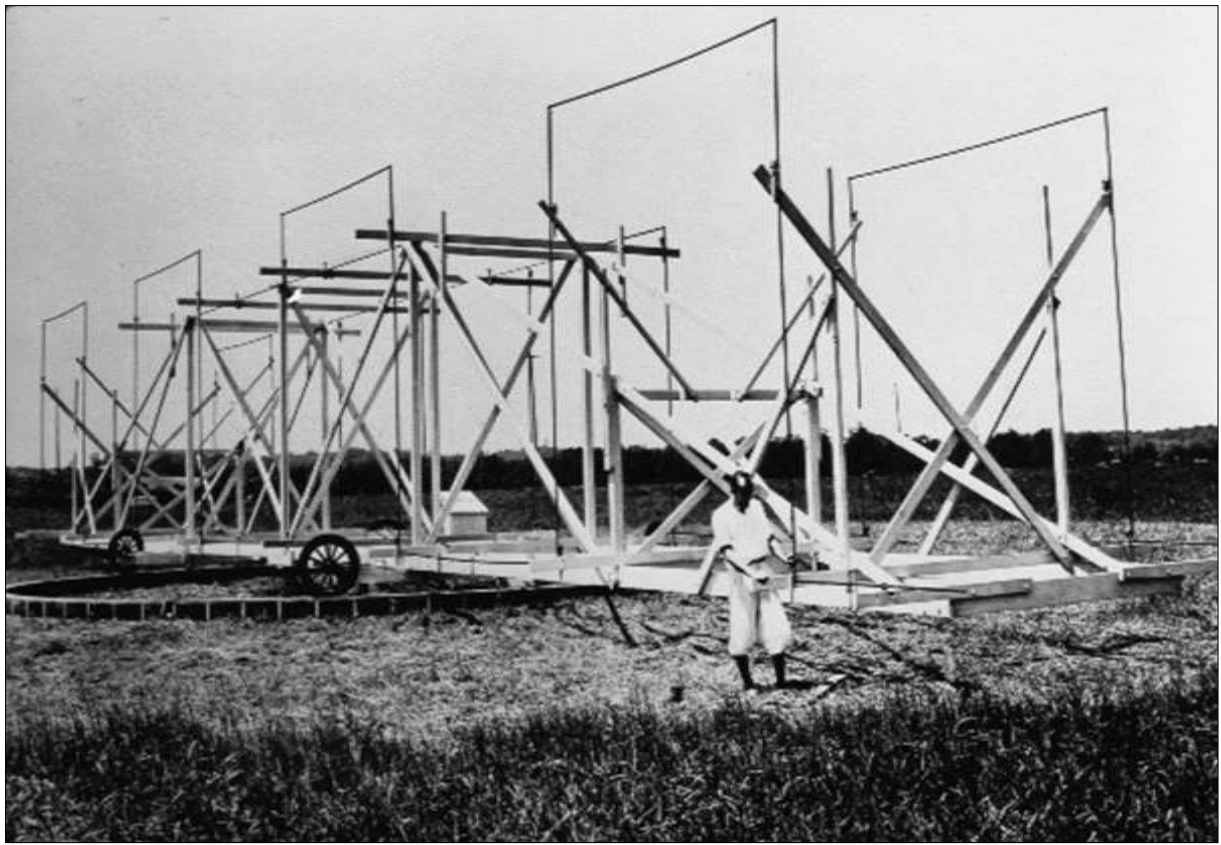

Figure 3: Karl Jansky at his rotating Bruce Array, near Holmdel, NJ (courtesy: Nokia Corporation and AT&T Archives).

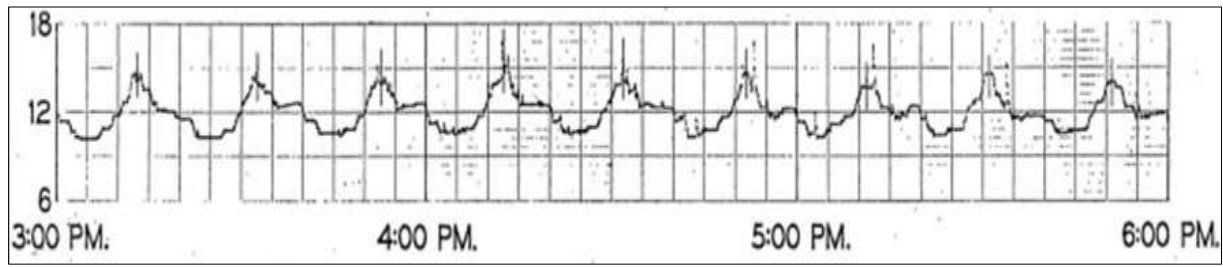

Figure 4: Reproduction of Jansky's 16 September 1932 chart recorder record showing the radio emission from the Milky Way peaking up three times an hour when his beam crossed the plane of the Milky Way (NRAO/AUI Archives).

Jansky's father was a Professor of Electrical Engineering and Karl regularly kept him informed about his work. So, he wrote to his father that "… it always seems to come from a direction that is the same or nearly the same as the direction the sun is from the antenna."[1] But then it seems that he was distracted by other telephone company work and he was not able to look at his data again until nearly a year later. And then he realized that the Sun had moved away from the direction of his radio noise. He again wrote to his father that the radio noise "… comes not from the sun, but from a direction fixed in space and I think is very startling."[2]

This *was* really startling—indeed exciting—radio noise from the Milky Way. This was 1933—space and astronomy weren't the same household words they are today. Nevertheless, with some prodding from his brother and the Bell Labs PR department, Jansky became famous. His story was carried by newspapers around the world including the front page of *New York Times*.[3] He was invited to give talks at professsional meetings and was interviewed on national radio including a live broadcast of what Jansky called Star Noise. He was only 28 years old and already world famous. But the reaction at Bell Labs was more subdued. This was 1933—probably the worst year of the Great Depression. Only a few months earlier Franklin Roosevelt had become the 32nd President of the US. Just a month earlier, he closed the banks. One-quarter to one-third of all US workers were unemployed. Millions of AT&T subscribers were forced to cancel their telephone subscriptions, the profits from which had supported research at Bell Labs, which was forced to lay off employees while others were cut to a 4-day work week. Galactic radio waves did not help build telephones, and Jansky spent most of the rest of his career working on telephone company business and then later on defense related electronics.

In several ways Jansky was lucky to have discovered cosmic radio noise. If he had been working at a longer wavelength, galactic radio noise would have been blocked by the Earth's ionosphere. If he worked at a shorter wavelength the galactic noise would have been too weak to observe with his primitive equipment. Also, Jansky was working the near minimum of the 11-year sunspot cycle. If he had done his experiments a few years earlier or a few years later when the Sun was more active, as it is now, the ionosphere probably would have blocked the cosmic radiation even at 15 meters.

For the following decade Grote Reber was the only person in the world to seriously follow up on Jansky's discovery. Reber took off the summer of 1937 from his job as a radio engineer and built a 32-foot dish in his backyard in Wheaton, Illinois. It was the world's first radio telescope. It was also by far the largest parabolic dish in the world, not exceeded until the Naval Research Laboratory erected a 50-foot dish in 1951. Reber was only 25 years old when he built his antenna and when he began a decade long study that brought radio astronomy to the attention of the astronomical community.

Reber was a good scientist. He knew about the Rayleigh-Jeans radiation law. He understood that if the galactic radio noise that Jansky had discovered came from a hot interstellar gas, the intensity should increase as the square of the observing frequency. So he designed his equipment to work at the highest frequency then feasible, 3,300 MHz. But he was surprised and disappointed that he did no detect anything. So, he redesigned his receiver to work at a lower frequency, 910 MHz. Still nothing. He built it again for 160 MHz, and finally detected Jansky's galactic radio noise, but it was about 100 times weaker than he expected (Figure 5). Reber understood that Jansky's star noise was not the kind of familiar thermal noise that increases with frequency, but that Jansky had discovered a new phenomenon—what we now call non-thermal radio emission—where the intensity decreases with increasing frequency. This strange frequency behavior was a complete surprise. It would be more than a decade before Jansky's star noise was understood to be coming from highly relativistic electrons moving in weak cosmic magnetic fields—or what we now call synchrotron radiation.

Now that he had his equipment working, Reber spent the next few years making the first detailed map of this non-thermal radio emission from the Milky Way (Reber,1944). He also had a regular daytime job in Chicago. Following a 1 hour commute each way by train to his job in Chicago, he would have dinner, observe until midnight, then he would sleep for a few hours and then go back work in Chicago.

By the late 1940's, Reber realized that he was no longer able to compete with the new well-funded university and government radio astronomy projects, so he so he decided to sell all of his equipment to the US Government and he went to work for the National Bureau of Standards (NBS) in Washington, DC. Reber's antenna was disassembled and it lay in storage for ten years in Northern Virginia at a site that is now part of Dulles Airport. In 1958 Reber brought all the pieces to Green Bank where he re-erected it in its current position at the entrance to the NRAO Green Bank Observatory. During his 1958 visit to Green Bank, Reber also initiated contact with Bell Labs to build a replica of the Jansky antenna which also now stands near the entrance to the Green Bank Observatory. One of Reber's later visits to Green Bank is shown in Figure 6.

## 3   A SEQUENCE OF SEREDIPITOUS DISCOVERIES CHANGES ASTRONOMY

When Reber was demonstrating his telescope to representatives from the NBS as part of his sales pitch, his chart recorder suddenly went off scale, and at the same time, everyone could hear a loud cracking noise coming from his receiver. What had happened was that he accidently discovered powerful radio waves from coming from the Sun that were so strong that they were coming in the sidelobes of his antenna even when it was not pointed at the Sun. And, he quickly dashed off a paper to *Nature* to stake his claim to the discovery of solar radio noise (Reber, 1946).

But this wasn't the first time anyone had detected radio noise from the Sun. Back in 1942, German planes were bombing Britain daily. Since they were expecting a full-scale German invasion, England built a chain of radar stations along the southern coast. However, on 12 February 1942, the German battleship *Scharnhorst*, along with several other German warships, managed to sail up the English Channel and were able to get safely back to their German port in Kiel. Because of intense interference, the German warships were undetected by the British radar.

This caused considerable concern, not only in England, but even in the US. Had Germany perhaps devised a new jamming technology that apparently made the radar defenses ineffective? There was an official parliamentary inquiry and calls for Winston Churchill to resign as Prime Minister.

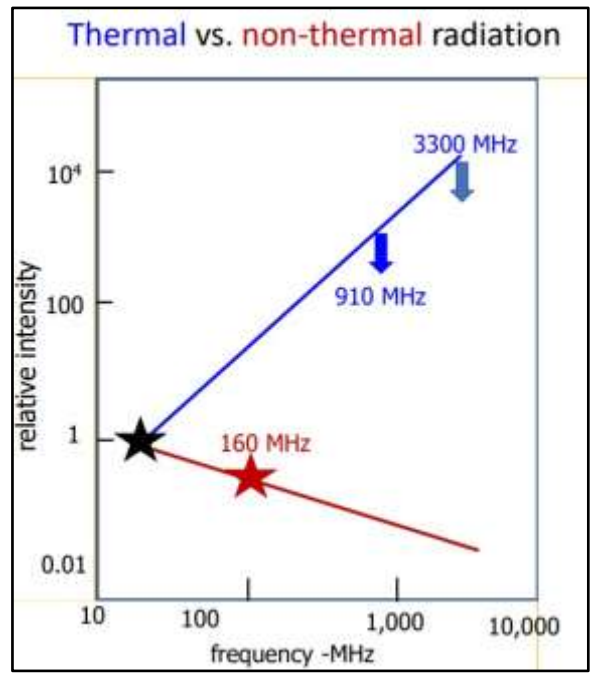

Figure 5: Expected frequency dependence of cosmic radiation based on the Rayleigh-Jeans radiation law (blue). Reber's negative results at 3300 MHz and then 910 MHZ are shown with arrows. Non-thermal spectrum (red) deduced by Reber based on his 160 MHz observation (red star star).

J. Stanley Hey was a civilian working for British Air Force. He was instructed to find out more about the apparent German radar jamming, which curiously seemed to continue even when there were no airplane or ship movements. He became suspicious when he realized that the apparent jamming occurred only in the daytime. He con-tacted the Royal Observatory and learned that on 12 February, there had indeed been intense solar activity, and he correctly concluded that the apparent radio jamming was coming, not from German transmitters, but from the Sun, and was associated with a giant sunspot that had appeared on 12 February.

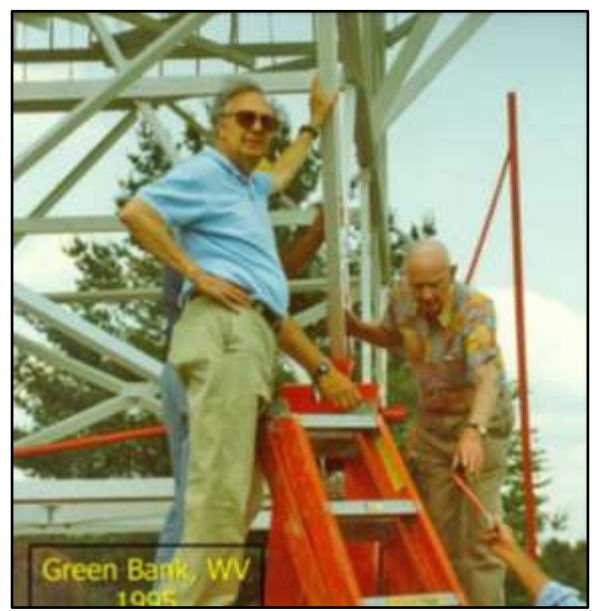

Figure 6: Grote Reber with the author at the time of his 1995 visit to Green Bank (Kellermann Collection).

Hey described his discovery in a detailed report.[4] But this was wartime, and his report was classified so Hey's discovery of powerful radio bursts from the Sun remained a secret until 1946. We now know that during the War many other radar operators, both Allied and German, apparently independently detected solar radio noise, but it was all kept secret. So, Grote Reber had the dubious honor of the first publication reporting strong radio bursts from the Sun (Reber, 1946).

After the war, Hey went on to make his own maps of the galactic radio noise, and he accidently discovered intense radio emission coming

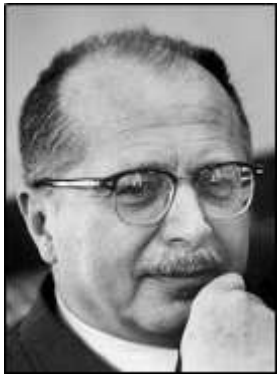

Figure 7: Caltech Professor Jesse Greenstein who missed out on the discovery of quasasr (courtesy: Caltech Archives).

1946). This was the first so called discrete radio source, Cygnus A. But there was no obvious optical counterpart that astronomers could associate with the radio source, and it was speculated it might be so far away that it was beyond the range of any optical telescope.

Shortly after that, John Bolton and Gordon Stanley in Australia discovered three other sources of strong radio emission. One of these radio sources lay in the direction of the Crab Nebula, which was well known to astronomers as the remains of a supernova that had exploded in the year 1054. The Crab Nebula is very far away and Bolton calculated it had to be an incredibly powerful radio source. They also found two other sources of radio noise in the same direction as two bright galaxies. One, NGC 5128 has the well-known dust band that crossed the galaxy and the other, M 87 has a jet like feature extending far away from a bright nucleus.

Bolton realized that if M87 and NGC 5128 were really galaxies, they would be more than a thousand times further away than the Crab Nebula. This meant that they would be more than a million times stronger than the Crab Nebula Apparently he could not comprehend that anything could be such a strong source of radio noise and so he argued that M87 and NGC 5128 were not really galaxies, but that they must be just gaseous nebula in our own Milky Way, and so they were really only as luminous as the Crab Nebula.

Nevertheless, their paper with the innocuous title, "Positions of Three Discrete Sources of Galactic Radio-Frequency Radiation" is considered by historians as the discovery of what we now all call 'radio galaxies', and John Bolton was recognized for his discovery of the first radio galaxies (Bolton et al., 1949).

Soon after this, astronomers were able to identify the radio source Cygnus A with a very faint and rather distant galaxy (Minkowski, 1960). Since Cygnus A is the strongest radio source in the sky, this suggested that the weaker radio sources were likely even further away and were probably far beyond the limit of even the largest optical telescopes. This generated a lot of new interest in radio astronomy which had previously been more or less ignored by what we can call the 'real' astronomers. Now the hunt was on to find the most distant radio galaxies. It was naturally assumed that further away the radio sources, the smaller they would appear. So, astronomers concentrated on observing small diameter radio sources.

It was around this time that John Bolton moved to Caltech where he started a new radio astronomy program. One of his goals was to find smaller and presumably more distant radio galaxies. Bolton and his colleagues were very successful and they did find many distant radio galaxies. But then there was an apparent setback. The radio source known as 3C 48 looked like a nearby star, not a galaxy. Using the Palomar 200-inch telescope, Allan Sandage discovered that 3C 48 was variable and changed brightness from hour to hour, characteristic of many stars, but certainly not of a galaxy. Moreover, 3C 48 had an unusual optical spectrum with very strange optical emission lines that no one could understand.

Jesse Greenstein (Figure 7) was a Professor at Caltech and an expert in optical spectroscopy. He spent several years trying to understand the spectrum of 3C 48. Meanwhile, John Bolton who had no training in spectroscopy, thought he could interpret the spectrum if one just assumed that the spectral lines were shifted by 37% which would mean that 3C 48 was really a distant galaxy in the expanding Universe, and was not just a nearby star.

However, Greenstein rejected Bolton's speculations, and essentially told him that as a radio astronomer, he did not understand optical spec-troscopy. Greenstein instead interpreted the spectrum in terms of exotic highly ionized rare earth elements that he suggested were located in the stellar atmosphere. He submitted a lengthy paper for publication in the *Astrophysical Journal* and handed out pre-prints to other Caltech professors. I was only a graduate student at the time, and did not merit a pre-print from Greenstein, but I was able to get an unauthorized copy from one of the other faculty members.

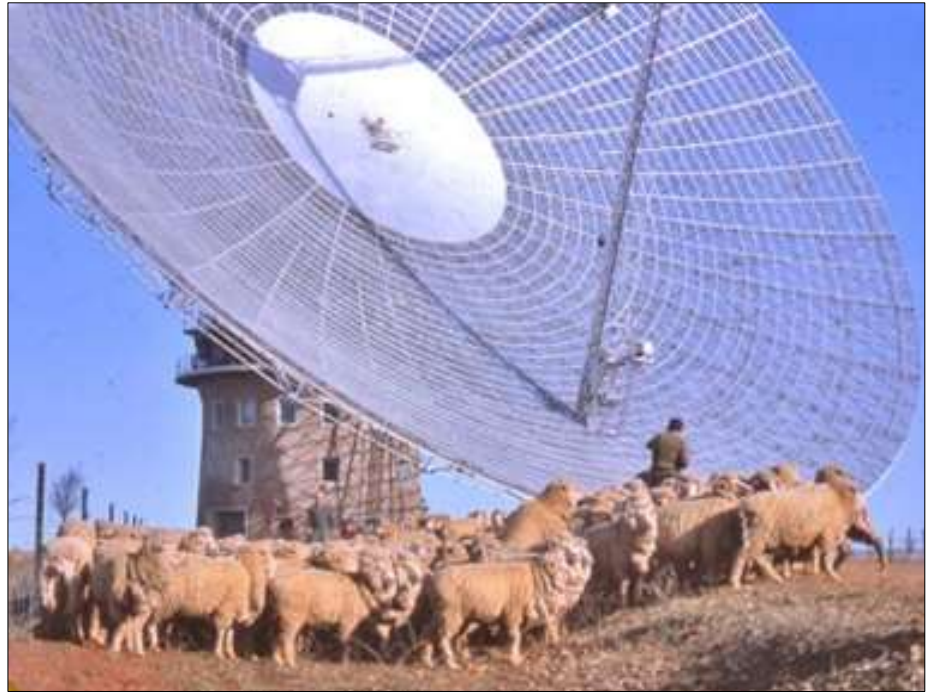

Figure 8: The Parkes Radio Telescope, which was used to determine the position of 3C 273 (Kellermann Collection).

The breakthrough came two years later when the strong radio source 3C 273 was occulted by the Moon. This is when the Moon crosses in front of the radio source and is similar to a solar eclipse. Since we know the position of the Moon very accurately, by timing when the radio source disappears behind the Moon, we can determine its accurate position on the sky. Astronomers had not had much interest to the radio source known as 3C 273 since it wasn't very small—so they thought it was a relatively nearby radio galaxy and sort of ignored it even though it was one of the brightest radio sources in the sky.

By this time John Bolton had moved back to Australia where he was now in charge of the new 64-m Parkes Radio Telescope (Figure 8). He took charge of the occultation, closed down the local radio station to prevent interference, and carved off a piece of the antenna so it could observe the whole occultation. The occultation data was used to determine a very precise position of 3C 273—good they thought to a second of arc. But they made a mistake in calculating the position of the Moon, so their position for 3C 273 was actually off by about 15 arcsec.

Bolton sent these incorrect coordinates to Maarten Schmidt at Caltech, and Schmidt was naturally confused. The optical image in the region of 3C 273 appeared to show both a bright star and a faint nebulosity. The position that Schmidt had received from Bolton, did not agree with either the star or the nebulosity. Schmidt assumed the nebulosity was a galaxy and probably the radio source, and that the star was just a random bright star in the image. It is not clear why Schmidt decided to observe the star. Perhaps it was just because it was a bright object near the radio source, or perhaps he was guided by the radio position determined by the Caltech radio astronomers that was close to that of the star.

But as with 3C 48, Schmidt found that the 3C 273 star had a very strange pattern of emission lines that somehow looked familiar to Schmidt, but yet they were not quite right (Figure 9). He puzzled over the spectrum for six weeks until he realized that the spectral lines in 3C 273 were just the familiar hydrogen lines—but they were shifted in wavelength by 16%. So, although 3C 273 looked like a nearby star, it had a large redshift which meant that it was located in a distant galaxy (Schmidt, 1963).

Schmidt became famous and appeared on the cover of *Time Magazine*. He had discovered a previously unknown, but major new constituent of the Universe. For the lack of any other name, Schmidt called these objects quasi-stellar radio sources, because they looked like stars but were not stars. This later got shortened to quasars. Armed with this new understanding of the spectra, Schmidt and Greenstein then went back to re-examine the spectrum of 3C 48, which they realized had an even larger redshift, 37%, or just the value that Bolton had claimed two years earlier and that Greenstein had rejected (Greenstein and Mathews, 1963).

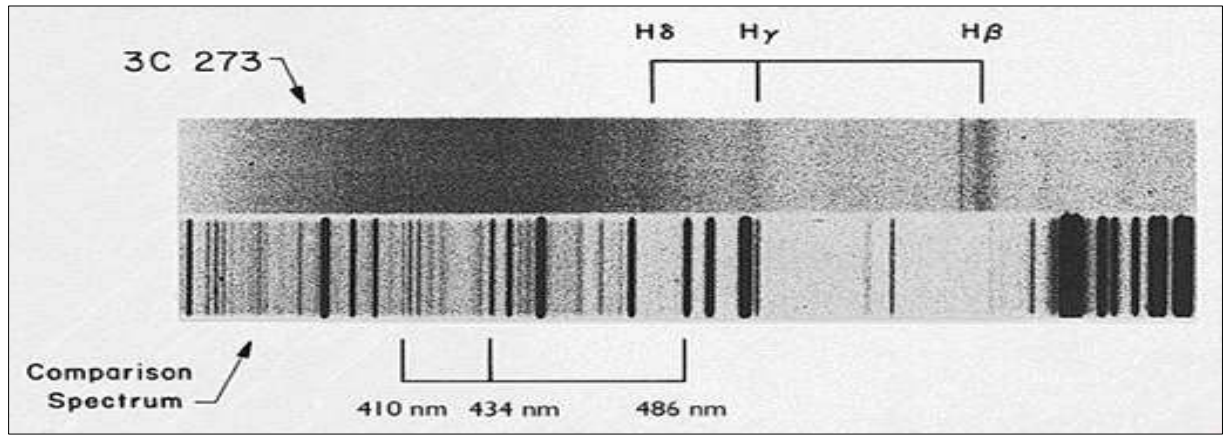

Figure 9:  Optical spectrum of 3C 273 (Credit:  Maarten Schmidt).

The identification of the bright radio source 3C 273 with the bright quasar was actually based on an accurate radio position measured with the Caltech interferometer, and not the Parkes occultation, but 3C 273 had been ignored until the Parkes lunar occultation, which initiated new interest in 3C 273, although the initial occultation position itself was incorrect (Hazard et al.,1963). We now recognize that quasars are very distant and are the most energetic objects in the Universe. As it later turned out, 3C 48, 3C 273, and many other quasars had been previously photographed by lots of astronomers as far back as 1880, but they were thought to be ordinary stars in our own Milky Way so they had been pretty much ignored.

After he realized that 3C 48 was not a star, and that his paper was all wrong, Greenstein was greatly embarrassed. He withdrew his paper from the *Astrophysical Journal* and he confiscated all the preprints that he had distributed. But since he did not know about my unauthorized copy, he did not get my preprint, so I think I have the only existing copy of what would have been a very embarrassing paper.

After he retired, I asked Greenstein why he had missed the discovery of quasars two years earlier by rejecting the possibility that 3C 48 had a large redshift. He told me, "I was known as a maverick, and I was afraid to go out on a limb."[5]

Since quasars are very bright, Schmidt's discovery initiated a furious new hunt to fid the most distant objects in the Universe. Tony Hewish was a radio astronomer in Cambridge, England. Margaret Clarke was one of his PhD students, and he assigned Margaret the task of finding new compact radio sources that might be distant quasars. Margaret became puzzled when some of her very small radio sources seemed to vary very rapidly—faster than once per second—but only when they were located near the Sun. She suspected that these rapid fluctuations were due to the irregularities in the solar wind that was rapidly flowing past the radio source and causing them to twinkle in the same way that stars twinkle due to irregularities in the Earth's atmosphere. The existence of a solar wind had actually been predicted much earlier, but Margaret Clarke accidently discovered the solar wind when she was looking for new quasars as part of her PhD project (Clarke, 1964).

Hewish was excited. Here was an opportunity to directly observe the solar wind. He built a new antenna to study this newly discovered phenomena. Actually, he did not build it himself. He used the time-honored tradition of having graduate students do the heavy labor of building the antenna array. One of these students, was Jocelyn Bell. When the array was finished, Jocelyn's job was to analyze the data which came out on long roles of paper charts—100 feet every day. She noticed a very peculiar radio source. Instead of the random twinkling that had been found by Margaret Clarke, Jocelyn's source appeared to be periodic, with one pulse about every 1.3 seconds.

Jocelyn reported her discovery to her supervisor. But Hewish was an experienced radio astronomer. He was confident that he knew the difference between terrestrial interference and a real cosmic radio source. So, he told Jocelyn that she was just seeing man-made interference and to get on with her work on the solar wind. But Jocelyn had already determined that the peculiar periodic radio source lay in a direction that was fixed in space, not fixed to a direction on the Earth. So, she told Hewish, "Well it may be man-made but not from an earth man who works on a 24 hour schedule not a 23 hour 56 minute [sidereal] schedule." (Bell Burnell, 1984).

This led to the idea that they might be seeing an interstellar beacon that extraterrestrial beings were using to navigate around the Galaxy. Somewhat whimsically, she labeled her radio source LGM 1 (Little Green Men 1). However, we now know that these periodic radio bursts come from pulsars, which are rapidly rotating highly magnetized stars, about 10 miles in diameter, made up of only neutrons packed very close together. The neutron star spins around about once a second, and ejects a beam of radiation. Every time that beam points toward the Earth, we see a pulse, like a lighthouse beacon.

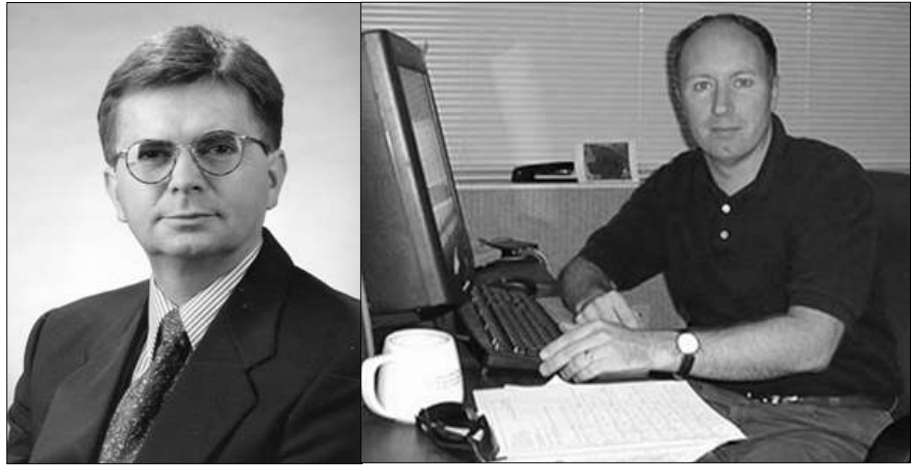

Figure 10: Alex Wolszczan (left) and Dale Frail (right) who discovered the first known planets outside our Solar System (courtesy: AIP Emil Segré Visual Archives).

Neutron stars were actually predicted back in the 1930s (Baade and Zwicky, 1934), but they were discovered accidently by Jocelyn Bell while she was studying something else, the solar wind, which in turn had been discovered by Margret Clarke when she was looking for more quasars. Apparently, the word pulsar was found on a Cambridge blackboard after a press conference. No one seems to know who came up with the word, but the name has stuck. Seven years later, Tony Hewish was recognized with the 1974 Nobel Physics Prize for his role in the discovery of pulsars. Jocelyn who had a hard time convincing Hewish that the pulses were not interference was not mentioned.

As it later turned out, around the time of the Cambridge work, the U.S. was operating the Ballistic Missile Early Warning System (BMEWS)—a chain of radar stations extending across North America from Alaska across northern Canada to Greenland. The BMEWS would give advanced warning of Russian missiles coming over the North Pole headed to the U.S. In 1967, Charles Schisler was the operator in charge of one of these radar stations located near Fairbanks, Alaska, and noticed that he appeared to be receiving radar echoes even when he wasn't transmitting. Schisler had been a navigator on a B-47 bomber so knew something about celestial navigation. He was able to work out the position in the sky where the pulses were coming from, and realized that they were coming from the direction of the Crab Nebula, not from Russian missiles. He then went on to locate a number of other pulsars.

Schisler found this pretty interesting, and after some library research realized that he had made an important scientific discovery (Schisler, 2008). But this was during the Cold War, so Schisler's remarkable discovery was classified and wasn't made public until 40 years later, decades after Joycelyn Bell's discovery and Hewish's Nobel Prize.

Some pulsars are actually pretty strong radio sources. The only reason that they had not been found earlier by radio telescopes, is because radio telescopes were not designed to detect pulsed radiation. Radar systems *are* designed to detect pulsed radiation, and probably many radar facilities had previously detected pulsars, but of course didn't recognize what they were.

The discovery of pulsars was very important and led to a number of other important discoveries. For example, in 1991, Alex Wolszczan was using the giant Arecibo radio telescope in Puerto Rico to observe pulsars. At the time, the Arecibo antenna was being repaired, so it could not be steered in the sky and he could only observe pulsars in a very narrow strip of the sky that passed directly overhead. One of the pulsars that fortunately lay in this small strip behaved strangely. Normally pulsars are very stable and are the most precise clocks known to man. But this one had curious variations in the pulse period.

After Dale Frail used the VLA to further study this pulsar, Wolszczan and Frail (1992) realized that the timing anomalies were caused by at least two planets that were orbiting the pulsar (Figure 10). The planets produced a small tug on the pulsar position which led to the timing irregularities. This was the first discovery of any planet that was located outside our own Solar System. Both planets had masses only a few times greater than that of the Earth. It was another two decades before any other Earth-like planets were discovered that were located beyond our Solar System.

Russell Hulse was still a graduate student when he was using the Arecibo radio telescope in the summer of 1973, and discovered a pulsar that was in orbit around another neutron star. Working with his University of Massachusetts Professor, Joe Taylor, they showed that the pulsar orbit was gradually changing as it was losing energy by gravitational radiation (Hulse and Taylor, 1975). Pulsars vary greatly in signal strength, from hour to hour and from day to day. Hulse was fortunate, in that on the day he discovered the double pulsar, it was only 3% stronger than his cutoff level. Had it been just a few percent weaker that day, he would have missed it.

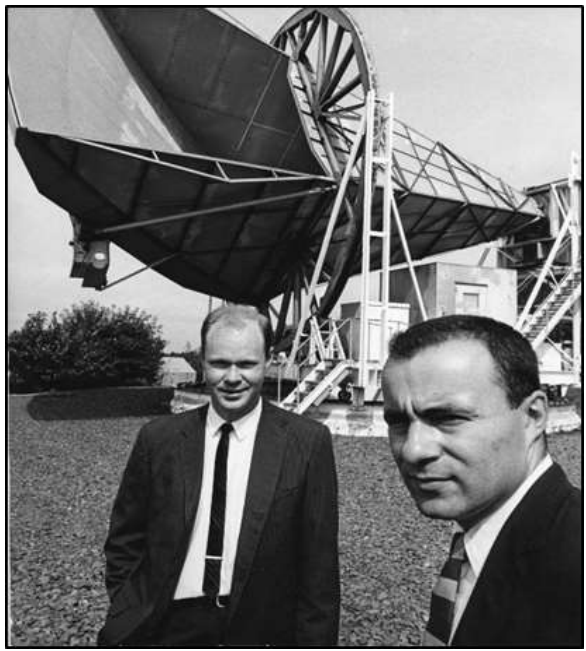

Figure 11: Robert Wilson and Arno Penzias near their Holmdel horn antenna (courtesy: Nokia Corporation and AT&T Archives).

The existence of gravitational radiation was predicted by Einstein (1916) back in 1916, but he thought that it would be so weak that it couldn't possibly ever be observed. In 1993, Hulse and Taylor shared the Nobel Prize in Physics for their discovery of the double pulsar and gravitational radiation.

Arguably, one of the most important astronomical events in the twentieth century was the discovery of the 3° cosmic microwave background radiation by Arno Penzias and Bob Wilson at Bell Labs. The cosmic background radiation is the weak radio glow coming from the whole sky that is left over from the original 'Big Bang' at the start of the Universe about 14 billion years ago.

Penzias and Wilson did not set out to discover the beginning of the Universe. Like Jansky they were working for the telephone company. In fact, their antenna was located very near the site where Jansky had built his antenna. Penzias and Wilson were using a very sophisticated 20-foot horn antenna that was originally designed to be used on telephone relay towers (Figure 11). They built what they thought was a very sensitive radio receiver, but they were discouraged when they found that there was more noise coming out of their antenna than they expected, so they assumed that they had missed some subtle instrumental effect.

While they were still trying to understand where their excess noise was coming from, Penzias accidently became aware of an experiment at nearby Princeton University that was in fact actually intended to detect the faint radio signal that might be left over from the Big Bang. After taking with the Princeton group, Penzias and Wilson realized that they had accidently discovered the faint radio noise that signaled the beginning of the Universe.

They published their discovery in a very short 1-page note in the *Astrophysical Journal* (Penzias and Wilson, 1965) that was almost entirely technical, and made no mention of any cosmological implications except for a one sentence reference to "… a possible explanation …" that was contained in a separate paper that was published by the Princeton group (Dicke et al., 1965).

Their one-page note became one of the most highly cited papers in astronomy. It has been referred to in about 1500 subsequent publications and has literally started a whole new field of cosmology. Hundreds if not thousands of scientists later became involved in studying the cosmic background radiation. It has also generated the construction of dozens of new radio telescopes, all intended to study the cosmic microwave background which was accidently discovered by Penzias and Wilson while they were testing their new receiver.

It later turned out there were many other scientists who had previously detected the cosmic microwave background, but they had either ignored it or swept it under the rug as an unexplained instrumental effect, or in several cases they just did not recognize its importance. But it was Penzias and Wilson who later received the 1978 Nobel Prize in Physics "For their discovery of the cosmic microwave background radiation."

## 4 REFLECTIONS ON THE PAST AND FUTURE OF RADIO ASTRONOMY

I have mentioned a number of Nobel Prizes won by radio astronomers. Actually, there is no Nobel Prize in Astronomy. Legend has it that this is be-because Nobel's wife ran off with an astronomer. In fact, there is no evidence that Nobel was ever married, but it makes a good story. Whatever the reason, until 1974 no one was ever recognized with a Nobel Prize for any work in any field of astronomy or astrophysics. But then, interestingly the first six scientists to receive a Nobel Prize for any astronomical discovery were all radio astronomers. By 2006, there were a total of eight radio astronomers who had received Nobel Prizes.

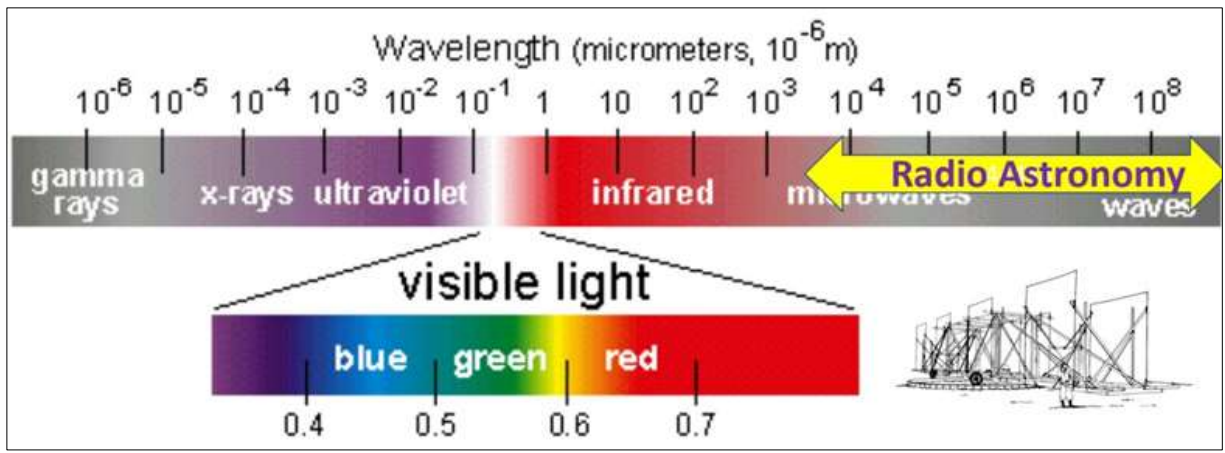

Figure 12: The electromagnetic Spectrum extending from gamma rays at the very short wavelength end to radio at the long wavelength end and showing the narrow slice of the spectrum available to astronomers before the discovery of cosmic radio waves.

Karl Jansky himself was nominated in 1948, but this was before the impact of his discovery was widely appreciated. Jansky died in 1950 when he was only 44 years old, so he was no longer eligible. Had he lived longer, I think surely Jansky would have received a Nobel Prize for his discovery of cosmic radio noise.

As I indicated at the start, following the pioneering work of Jansky and Reber, there were many other discoveries by radio astronomers that completely changed the landscape of astronomy but I didn't have time to discuss all of them. These include, electrical storms on Jupiter, interstellar molecules, cosmic masers, apparent faster than light motion in quasars, gravitational lenses, and many other things.

So, why were radio astronomers the first to discover all these phenomena that were previously unknown? Remember, astronomy is an observational science. Astronomers cannot do experiments. All we can do is look. Way back in 1609 Galileo used a telescope for the first time to study astronomical bodies. He unexpectedly or serendipitously discovered the moons of Jupiter and the phases of Venus. For the next three hundred years astronomers built bigger and better telescopes, but they were all confined to the narrow slice of the electromagnetic spectrum known as visible light (Figure 12). That all changed with Karl Jansky's discovery of radio noise from the Milky Way. For the first time astronomers could observe outside the traditional visible or optical 'window'. More recently, orbiting observatories have opened up other parts of the electromagnetic spectrum: X rays, gamma rays, the infrared, and the ultra violet, which are all blocked by the Earth's atmosphere, are all now part of everyday astronomy. But radio astronomy was the first to reach out beyond the visible spectrum, so it is not surprising that it was radio astronomers who were the first to uncover all these previously unknown phenomena.

In Figure 13 I have plotted the number of radio astronomy discoveries as a function of time. As you can see, it started out slow with Jansky and Reber. Then the number of discoveries rapidly rose during the 1960s and 1970s, but then leveled off with only a few new discoveries made during the past 40 years. This was very different from the first 40 years. What happened?

Well, maybe by 1985 we had discovered everything, and there was nothing left to discover. Probably this is not right. Going back to when the world was supported by four elephants standing on the backs of a turtle, man has never understood the Universe. There is no reason to believe we know it all now. But perhaps we have 'gathered the low-hanging fruit', and the way forward is going to be more difficult and slower.

Maybe the explanation is we have run out of new technologies, or maybe our instruments are just getting too big and too expensive to afford. Also, the data from modern radio telescopes are so extensive and so sophisticated that the scientists don't get to see their data until it has been partially digested by a computer. As Jocelyn Bell (1968) wrote in her thesis:

> If the aerial output had been digitized and fed directly to a computer [pulsars] might well have not been discovered because the computer would not have been programmed to search for unexpected objects.

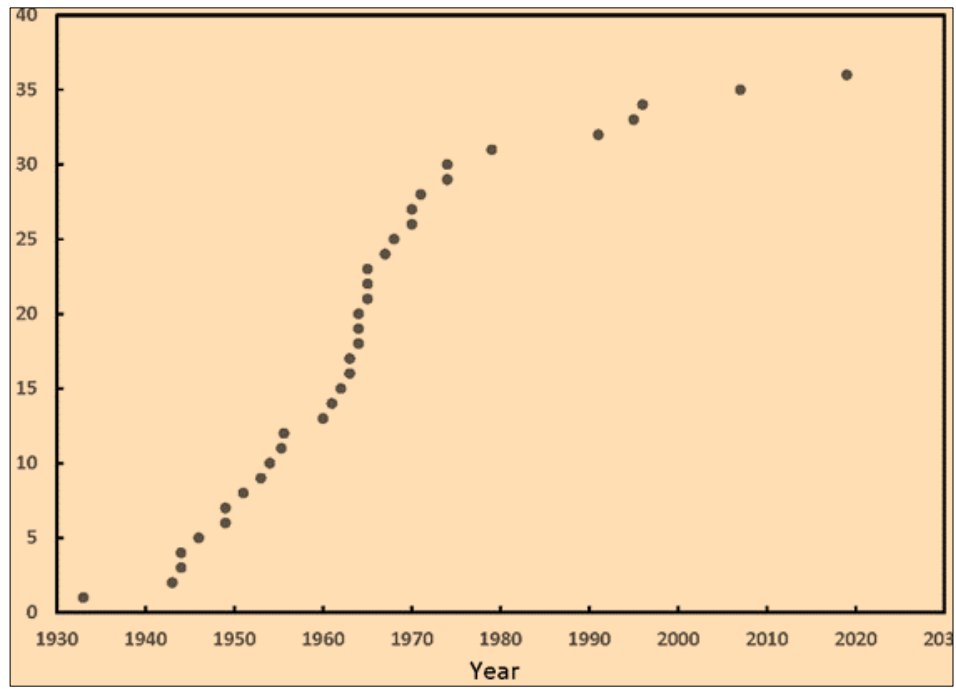

Figure 13:   Plot showing the cumulative number of radio astronomy discoveries as a function of time since the early 1930s (plot: the author).

Perhaps the next step will be to use Artificial Intelligence to analyze our data. Maybe this will lead to new discoveries or maybe it will deter new discoveries. I do not know. Also, over the last half century or so, the sociology of science has changed a lot. In the pioneering days of radio astronomy, a single individual, or perhaps a small team of two or three individuals, could conceive a program, design and build the equipment, carry out the observations, analyze and then interpret their data. They understood every aspect of what they were doing, and were able to distinguish between subtle instrumental effects and genuine new discoveries.

Today, the highly skilled engineers who design and build our instruments do not use them, and the equally skilled astronomers that use them don't build them and so may not fully understand their limitations. Also, I suspect that some of our instruments are now so complex that there is no single individual who fully understands the whole instrument.

Another change is peer review. It has long been traditional for scientific journals to have papers reviewed by experts before publication. The idea was to be sure no nonsense gets published. But, now, the competition to use our large telescopes is very intense and there are typically many more proposals to use the telescopes than available time. So, proposals to use large telescopes are now evaluated by a peer review process. In other words, other astronomers get to tell us what we can and cannot study.

To exacerbate the problem, it is very expensive to build and operate large radio telescopes, so there is great pressure for each project to produce results. Therefore, the peer review process tends to be rather conservative and avoids taking risks. This means, that in order to get permission to use a telescope, or even to get a job or a research grant, scientists are asked to describe their expected results in such detail that, if it is known so well in advance what they were going to learn, the proposed investigation might not even be necessary.

It was not always like this. The original VLA proposal was motivated by the need for better sensitivity, resolution, versatility, and speed. The scientific justification was based entirely on obtaining pictures of radio galaxies and quasars. The proposal was only for continuum studies, and stated that "The addition of line spectrometer equipment … appears to increase the complexity of the system to a point not constant with the present state of the electronic art."[6] This raised much criticism from the molecule hunters; nevertheless, the final proposal only added "The selection of particular sub-systems should not preclude the ultimate use of the equipment for line work."[7]

Tommy Gold was a well-known astrophysicist but was also known as a very outspoken scientist. He liked to compare astronomers to sheep and claimed that unless one followed the herd they might be ridiculed or ignored. So, he argued that there is a lot of pressure for scientists to follow the herd. Especially young scientists. Gold (1984) further argued that following the herd, may be a good tactic for advancing one's scientific career but probably not good for the progress of science.

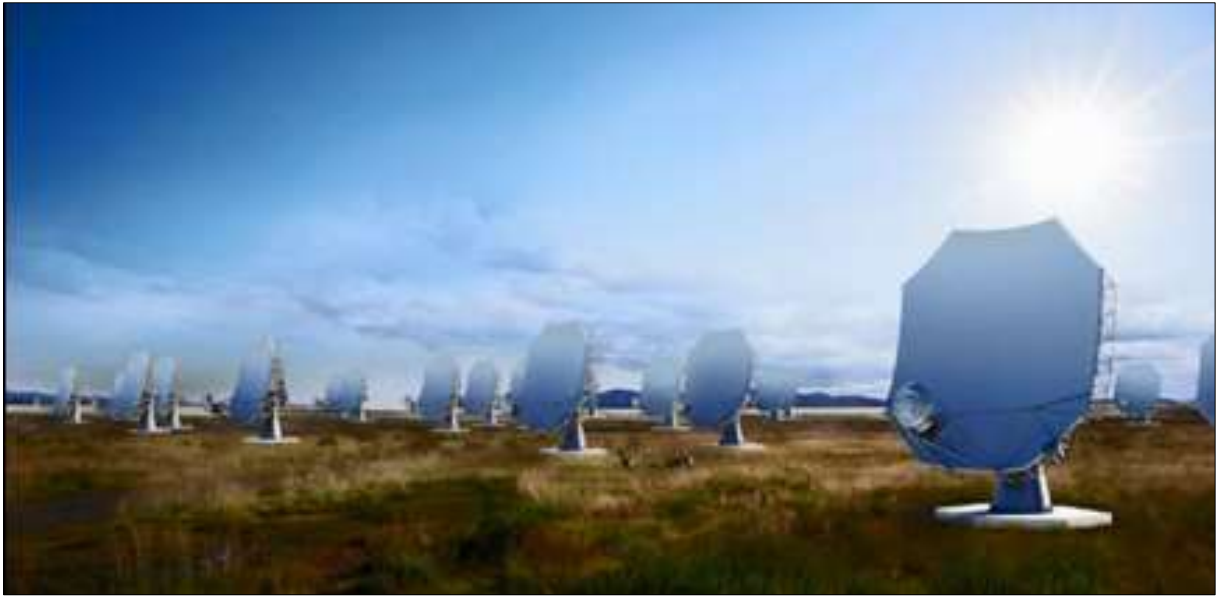

Figure 14: Artist's conception of the next generation VLA, which will consist of an array of 263 parabolic dishes spread over nearly 9 thousand kilometers (courtesy: NRAO/AUI/NSF).

In 2002, former Secretary of Defense Donald Rumsfeld, was asked at a press conference about Iraq Weapons of mass destruction. He explained:

> There are *Known Knowns*: There are things we know we know. We also know that there are *Known Unknowns*. That is to say there are some things we know we do not know. But there are also *Unknown Unknowns*; the ones we don't know we don't know.

So, what has this got to do with radio astronomy? The *Known Knowns* are things we think we understand. The things I have been talking about for the past hour, and the many other things that we think we now understand. The *Known Un-knowns* are things we know we do not know, like the rate of expansion of the Universe, the nature of dark matter, dark energy, and Fast Radio Bursts. Or, Is there other intelligent life in the Universe?

Most radio telescopes operating today are doing research on topics that were not even anticipated when the telescope was proposed. As Peter Wilkinson (2004) has pointed out, "Radio telescopes are not known for what they were built for!" For example, the original VLA proposal did not consider observations of stars or pulsars or molecular spectroscopy. And, of course, Fast Radio Bursts were not even mentioned when CHIME was proposed to study hydrogen in the early Universe.

So, if the history of radio astronomy is any example, the excitement of the next generation of radio telescopes such as the ngVLA (Figure 14) will not be in the old questions which they answer but in the new questions that will be raised by the new observations. They are the *Unknown Unknowns.*

## 5  NOTES

1. Karl Jansky to family, 18 January, 1932.
2. Karl Jansky to family, 10 June 1933.
3. New York Times, 5 May 1933.
4. Hey's March 1942 classified report, "Notes on G.L. [Gun Laying Radar] 27 and 28 February" AVIA7/3544 is filed at the U.K. Public Records Office in Kew, England.
5. Jesse Greenstein, private communication to the author, January 1995.
6. Staff of NRAO, 1965. *VLA Report Number 1.* 10 December. Green Bank, NRAO/AUI.
7. NRAO, 1967. The VLA. *A Proposal for a Very Large Array Radio Telescope.* Vol. 1. January 1967. NRAO/AUI.

## 6  ACKNOWLEDGEMENTS

My 2025 Jansky Lectures were largely based on material contained in the book, *Star Noise: Discovering the Radio Universe* written by Kenneth Kellermann and Ellen Bouton (2023). I am indebted to Ellen for her long-time collaboration and in particular for her thoughtful and constructive comments on an early draft of this man-uscript. Unless otherwise credited, illustrations are from the NRAO/AUI Archives. The NRAO is operated by Associated Universities Inc. under Cooperative Agreement with the National Science Foundation.

## 7 REFERENCES

Baade, W., and Zwicky, F., 1934. Remarks on super-novae and cosmic rays. *Physical Review*, 46, 76–77.

Bell, S.J., 1968. *The Measurement of Radio Source Diameters Using a Diffraction Method.* PhD Thesis, University of Cambridge, Cambridge, England.

Bell Burnell, S.J., 1984. The discovery of Pulsars. In Kellermann, K.I., and Sheets, B. (eds.), *Serendipitous Discoveries in Radio Astronomy*. Green Bank, NRAO/AUI. Pp. 160–170.

Bolton, J.G., Stanley, G.J., and Slee, O.B., 1949. Positions of three discrete sources of Galactic radio-frequency radiation. *Nature*, 164, 101–102.

Chu, S., 2015. Restoring the Foundation: Reviving the U.S. Science, Engineering, and Technology Enterprise, *Bulletin of the American Academy of Arts and Science*, Summer Issue 18.

Clarke, M.E., 1964a. *Two Topics in Radiophysics.* PhD Thesis, University of Cambridge, Cambridge, England.

Dicke, R.H., Peebles, P.J.E., Roll, P.G., and Wilkinson, D.T., 1965. Cosmic back-body radiation. *Astrophysical Journal*, 142, 414–419.

Einstein, A., 1916. The foundation of the General Theory of Relativity. *Annalen der Physik*, 49, 769–822 (in German).

Gold, T., 1984. The Reception of New Ideas by the Scientific Community. Paper presented at the Third Annual Meeting of the Society for Scientific Exploration, October 1984.

Greenstein, J.L., and Matthews, T.A., 1963. Redshift of the radio source 3C 48. *Nature*, 197, 1041–1042.

Hazard, C., Mackey, M.B., and Shimmins, A.J., 1963. Investigation of the radio source 3C 273 by the method of lunar occultations. *Nature*, 197, 1037–1039.

Hey, J.S., Parsons, S.J., and Phillips, J.W., 1946. Fluctuations in cosmic radiation at radio frequencies. *Nature*, 158, 234.

Hulse, R.A., and Taylor, J H., 1975. Discovery of a pulsar in a binary system. *Astrophysical Journal*, 195, L51–L53.

Kellermann K.I., and Bouton, E.N., 2023. *Star Noise: Discovering the Radio Universe*. Cambridge, Cambridge University Press.

Minkowski, R., 1960. A new distant cluster of galaxies. *Astrophysical Journal*, 132, 908–910.

Penzias, A.A., and Wilson, R.W., 1965. A measurement of excess antenna temperature at 4080 Mc/s. *Astrophysical Journal,* 142, 419–421.

Reber, G., 1944. Cosmic static. *Astrophysical Journal*, 100, 279–287.

Reber, G. 1946. Solar radiation at 480 Mc./sec. *Nature*, 158, 945.

Schisler, C., 2008. An independent 1967 discovery of pulsars. In Bassa, C.G., Wang, Z., Cumming, A., and Kaspi, V. (ed.), *40 Years of Pulsars: Millisecond Pulsars, Magnetars, and More.* New York, American Institute of Physics. Pp. 642–645.

Schmidt, M., 1963. 3C 273: a star-like object with large redshift. *Nature*, 197, 1040.

Wilkinson, P.N., Kellermann, K.I., Ekers, R.D., Cordes, J.M., Lazio, W., and Joseph, T., 2004. The exploration of the unknown. *New Astronomy Reviews*, 48, 1551–1563.

Wolszczan, A., and Frail, D.A., 1992. A planetary system around the millisecond pulsar PSR 1257+12. *Nature*, 355, 145–147.

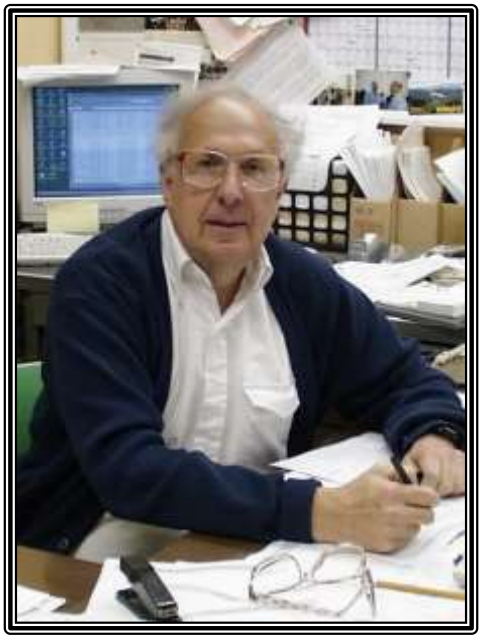

**Dr. Ken Kellermann** is a Senior Scientist, Emeritus, at the National Radio Astronomy Observatory (NRAO) in Charlottesville, Virginia. He was born in New York City in 1937 and received his SB in Physics from the Massachusetts Institute of Technology in 1959 and his PhD in Physics and Astronomy from the California Institute of Technology in 1963. Following a two year postdoctoral appointment at the CSIRO Radiophysics Laboratory in Sydney, Australia, Kellermann joined the NRAO. Except for two years spent as a Director of the Max Planck Institute for Radio Astronomy in Bonn, Germany, he has spent most of his career at NRAO where he has served as Acting Assistant Director for Green Bank Operations, Chief Scientist, and Head of the New Initiatives Office. He has also held adjunct appointments at the University of Virginia, the University of Arizona, and the University of Pennsylvania as well as visiting appointments at the CSIRO Radiophysics Laboratory, Caltech, and Leiden.

Kellermann is a member of the IAU, URSI, and the AAS. He is the past Chair of the IAU Commission on Radio Astronomy, and past Chair and current Executive Committee member of the IAU/URSI Working Group on Historical Radio Astronomy. He is also a Fellow of the American Academy of Arts and Science and the American Philosophical Society, a Member of the National Academy of Sciences, a Foreign Member of the Russian Academy of Science, and an External Member of the German Max Planck Society. He was the recipient of the AAS Helen B. Warner Prize, the NAS Gould Prize, the ASP Bruce Medal, and shared the 1971 AAAS Rumford Medal.

Kellermann's research has been largely devoted to the study of the radio spectra, time variability, and morphology of radio galaxies and quasars along with the techniques of Very Long Baseline Interferometry. More recently, he has concentrated on the history of radio astronomy.

Kellermann is the author or coauthor of *Galactic and Extragalactic Radio Astronomy* (with Verschuur), *Open Skies: The National Radio Astronomy Observatory and its Impact on US Radio Astronomy* (with Bouton and Brandt), *Star Noise: Discovering the Radio Universe* (with Bouton), and is the editor of the English language edition of *A Brief History of Radio Astronomy in the USSR*.